\documentclass[
twocolumn,
english,
superscriptaddress,
amsmath,
floatfix,
longbibliography,
floats,
prl,
]{revtex4-2}

\usepackage[T1]{fontenc}
\usepackage{lmodern}     
\usepackage{microtype}   
\usepackage{bm}

\usepackage{mathtools}   
\usepackage{amssymb}
\usepackage{braket}
\usepackage{mathrsfs}    

\usepackage{graphicx}
\usepackage[table]{xcolor}
\usepackage{array}
\usepackage{booktabs}     
\usepackage{multirow}
\usepackage{diagbox}
\usepackage{dcolumn}      
\usepackage{makecell}
\usepackage{siunitx}

\usepackage{float}
\usepackage{indentfirst}
\usepackage[normalem]{ulem} 
\usepackage{ragged2e}
\usepackage{enumitem}     

\usepackage{algorithm}      
\usepackage{algpseudocode}  

\usepackage[colorlinks=true,linkcolor=magenta,citecolor=magenta,urlcolor=magenta]{hyperref}

\newcommand{\OU}{Graduate School of Engineering Science, The University of Osaka, 1-3 Machikaneyama, Toyonaka, Osaka 560-8531, Japan}
\newcommand{\FI}{Center for Computational Quantum Physics, Flatiron Institute, New York, NY 10010, USA}
\newcommand{\BU}{Department of Physics, Boston University, Boston, Massachusetts 02215, USA}

\begin{document}

\title{Tensor network surrogate models for variational quantum computation}

\author{Ryo Watanabe}
\affiliation{\OU}

\author{Dries Sels}
\affiliation{\BU}
\affiliation{\FI}

\author{Joseph Tindall}
\affiliation{\FI}

\begin{abstract}
We adopt a two-dimensional tensor-network (TN) ansatz to simulate variational quantum algorithms on two-dimensional qubit architectures, demonstrating its capability to accurately simulate deep circuits through the Quantum Approximate Optimization Algorithm (QAOA) applied to Ising spin-glass problems on heavy-hexagonal and square lattices.
For heavy-hexagonal problems with up to three-body interactions, parameters trained on small instances and transferred to systems an order of magnitude larger improve the sampled energy distribution only up to intermediate depths, indicating a fundamental limit of parameter concentration as a transfer strategy.
By extending the training itself with TN simulations on larger system sizes, we avoid local minima and obtain lower-energy samples.
Analyses of entanglement growth and importance sampling show that the simulation remains classically feasible with moderate bond dimension.
We find that parameter concentration also persists on square lattices, albeit at substantially higher computational cost to perform reliable sampling.
Overall, our TN framework not only provides an efficient and controlled framework for benchmarking variational quantum algorithms on two-dimensional lattices, but also serves as an effective surrogate model for training variational algorithms.
\end{abstract}

\maketitle

\section{Introduction}

Quantum computing is a promising approach for tackling problems for which no efficient classical algorithms are known, including integer factorization and the simulation of quantum many-body systems~\cite{Shor1997,Feynman1982,Lloyd1996}.
This promise stems fundamentally from its ability to exploit uniquely quantum phenomena.
Current quantum devices, however, are inherently error-prone, and fully fault-tolerant quantum computation based on quantum error correction codes~\cite{PhysRevA.52.R2493} remains beyond the state of the art.
Consequently, considerable effort is devoted to identifying tasks for which currently available quantum hardware can offer practical value beyond existing classical methods.

A large body of literature has recently focused on demonstrating the notion of quantum ``advantage'' or ``utility''~\cite{Kim2023,King2023,doi:10.1126/science.ado6285,doi:10.1126/sciadv.adu9991}, creating a ``tug-of-war'' between quantum and classical methods over which approach is favorable~\cite{PRXQuantum.5.010308,PhysRevResearch.6.013326,Begu_i_2024,tindall2025dynamicsdisorderedquantumsystems}.
It is reasonable to take the stance, however, that the practical value of quantum hardware will ultimately be judged by whether executed circuits and their resulting measurements can clearly resolve scientifically meaningful questions that remained beyond the reach of classical methods~\cite{hasselgren2025probabilisticcomputingoptimizationcomplex,Abanin2025}.

Based on this criterion, one of the most promising and viable candidates for current hardware is the quantum approximate optimization algorithm (QAOA)~\cite{farhi2014quantumapproximateoptimizationalgorithm}.
QAOA is a representative instance of a variational quantum algorithm (VQA), combining parameterized quantum circuits with classical optimization procedures~\cite{cerezoVariationalQuantumAlgorithms2021} for solving combinatorial optimization problems.
While most applications of VQA aim to evaluate expectation values through many repeated measurements of the same observable~\cite{Peruzzo2014, PhysRevA.98.032309}, in QAOA one can extract solutions by sampling bitstrings from the output state~\cite{doi:10.1126/science.1057726}.
Recent studies have shown that classical surrogates can approximately simulate noisy circuits since quantum noise degrades the correlations, and observables can be evaluated efficiently within these approximate states~\cite{PRXQuantum.5.010308,10.1063/5.0269149,Begu_i_2024,rudolph2023classicalsurrogatesimulationquantum,PhysRevLett.133.120603,bermejo2026quantumconvolutionalneuralnetworks}.
In contrast, extracting solutions by sampling is believed to still remain extremely challenging for classical methods, even when the state itself is approximately representable~\cite{aaronson2010computationalcomplexitylinearoptics, Lund2017, Arute2019, Zlokapa2023, Morvan_2024}.
Accordingly, a series of hardware demonstrations has recently evaluated QAOA on superconducting platforms. Device noise, however, has so far limited these experiments to shallow circuits~\cite{QA_vs_QAOA_127, Short_Depth_QAOA_QA}.
This motivates accurate classical surrogates that can probe deep schedules and their behavior at scales beyond current quantum hardware.

In this work, we achieve this. We employ tensor-network (TN) methods to simulate QAOA, focusing on two-dimensional lattices native to modern superconducting quantum processors.
For parameter optimization, we rely on parameter concentration~\cite{Akshay_2021} --- whereby parameters trained on representative instances of a given class can generalize to other instances in the same class --- to set QAOA angles.
Specifically, we train parameters on small instances, and transfer these parameters into larger instances.
We also adopt a low-dimensional interpolation scheme that facilitates their extension to deeper circuit schedules~\cite{apte2025iterativeinterpolationschedulesquantum}.
We simulate QAOA circuits with a connectivity-aware TN ansatz, performing the necessary truncations conditioned on a belief-propagation (BP) approximation~\cite{10.21468/SciPostPhys.15.6.222,PhysRevResearch.3.023073}. We then \emph{approximately} sample bitstrings from the resulting states using boundary-MPS techniques adapted to arbitrary planar topologies~\cite{PhysRevB.104.235141, rudolph2025simulatingsamplingquantumcircuits}.

We benchmark our approach on Ising spin-glass problems defined on both heavy-hexagonal~\cite{QA_vs_QAOA_127, Short_Depth_QAOA_QA} and square lattice geometries.
For heavy-hexagonal instances, we investigate how system size and simulator fidelity during training influence parameter transfer.
Across a range of training system sizes, we observe a consistent behavior: transfer quality improves as the schedule depth increases, up to a saturation point, beyond which additional layers yield no further improvements.
We analyze the entanglement growth during circuit evolutions, and we find that it remains bounded, indicating that the resulting QAOA states remain classically tractable within our TN approximation.
Importance weights obtained by using boundary-MPS techniques, which quantify the sampling accuracy, improve systematically with the boundary MPS rank, while the sampled-energy histograms remain stable, suggesting that accurate sampling is already achieved at moderate boundary MPS bond dimension.

Comparing different training sizes, we find that optimizing on more representative systems, whose sizes more closely match those of the target instances, is most effective for parameter transfer; it helps the optimization escape local minima and produces energy distributions that converge to lower values.
To obtain these results, we explicitly make use of TN simulations for parameter optimization since these systems are already beyond the reach of state-vector simulations.
We thereby demonstrate that TN simulators are practical tools not only for forward simulations but also for parameter optimizations.

For square-lattice instances, where the BP approximation is less reliable due to the high density of short loops, we confirm that the parameter concentration remains effective and that the TN-based framework remains applicable with appropriate choices of the bond dimensions. Accurate sampling does, however, require significantly larger computational resources compared to the heavy-hexagonal case.
Collectively, our benchmarks indicate that a connectivity-aware TN ansatz allows for simulating deep-schedule QAOA circuits defined on two-dimensional lattices and that it can serve as an effective surrogate model for training VQA parameters for system sizes beyond the reach of exact diagonalization.

The remainder of this paper is organized as follows.
Section~\ref{sec:background} introduces QAOA for combinatorial optimization problems (Sec.~\ref{subsec:qaoa}), and defines the target problem class (Sec.~\ref{subsec:problem_definition}).
Section~\ref{sec:method} describes the parameter optimization strategy (Sec.~\ref{subsec:parameter_optimization}), and the TN techniques used for circuit simulations and sampling (Sec.~\ref{subsec:tn_methods}).
Section~\ref{sec:results} presents our numerical results, and Sec.~\ref{sec:conclusion} concludes with an outlook.
Supporting details are provided in the Appendices.

\section{Background}\label{sec:background}

\subsection{Quantum Approximate Optimization Algorithm}\label{subsec:qaoa}

QAOA attempts to find the minimum value of a cost function $C(\bm{z}): \bm{z}\in\{ \pm1 \}^n \rightarrow \mathbb{R}$ for a given instance of a combinatorial optimization problem as follows~\cite{farhi2014quantumapproximateoptimizationalgorithm}.
First, we write $C(\bm{z})$ as a classical Hamiltonian $H_C$ defined by $\bra{\bm{z}} H_C \ket{\bm{z}} = C(\bm{z})$, acting on $n$ qubits in the computational basis, and introduce the mixing Hamiltonian $H_X = \sum^{n}_i X_i$, where $X_i$ is the $X$ Pauli operator acting on the $i$-th qubit.
Then QAOA quantum circuits consist of a sequence of parameterized unitary gates:
\begin{equation}\label{eq:qaoa_ansatz}
    U({\bm{\gamma}, \bm{\beta}}) = \prod^p_{j=1} e^{-i\beta_jH_X}e^{-i\gamma_jH_C}~,
\end{equation}
with parameters $\bm{\gamma} = (\gamma_1, \cdots, \gamma_p)$ and $\bm{\beta} = (\beta_1, \cdots, \beta_p)$, where $p$ is the number of layers.
In order to minimize $C(\bm{z})$, we optimize the parameters $(\bm{\gamma}, \bm{\beta})$ such that the expectation value $\bra{\bm{\gamma}, \bm{\beta}} H_C \ket{\bm{\gamma}, \bm{\beta}}$ is minimized, where $\ket{\bm{\gamma}, \bm{\beta}} = U({\bm{\gamma}, \bm{\beta}})\ket{+}^{\otimes n}$.
We then draw bitstrings $\bm{x}$ from $\ket{\bm{\gamma}, \bm{\beta}}$, where $\bm{x} \sim P(\bm{x}) = | \braket{\bm{x} \vert \bm{\gamma}, \bm{\beta}} |^2$ is the probability of measuring $\bm{x}$ in the output state $\ket{\bm{\gamma}, \bm{\beta}}$.

\subsection{Problem definition}\label{subsec:problem_definition}

We consider Ising spin-glass problems defined on two types of lattices: heavy-hexagonal and square-grid lattices.
The heavy-hexagonal lattice is IBM's native architecture for superconducting quantum processors, and the corresponding problem was originally introduced in Refs.~\cite{QA_vs_QAOA_127, Short_Depth_QAOA_QA, pelofskeScalingWholechipQAOA2024, sachdeva2024quantumoptimizationusing127qubit} to benchmark QAOA on IBM's heavy-hex hardware.
The square-grid architecture is a dominant design for scaling superconducting processors nowadays as it naturally implements two-dimensional surface codes~\cite{Acharya2025}.

\begin{figure}[t]
    \centering
    \includegraphics[clip, width=3.0in]{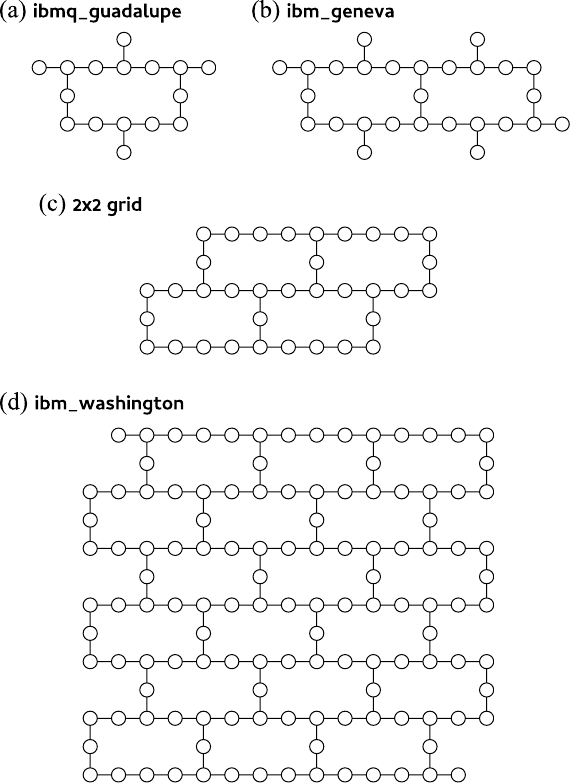}
    \caption{
        Schematic diagrams of heavy-hexagonal lattices considered in this work: (a) \texttt{ibmq\_guadalupe} ($n = 16$), (b) \texttt{ibm\_geneva} ($n = 27$), (c) $2 \times 2$ grid system ($n = 35$), and (d) \texttt{ibm\_washington} ($n = 127$).
        (a), (b), and (d) correspond to existing architectures of IBM quantum processors.
    }
    \label{fig:heavyhex_lattices}
\end{figure}

Let the heavy-hexagonal lattice be defined as $G = (V, E)$ with a vertex set $V$ and an edge set $E$; representative heavy-hexagonal lattices are schematically illustrated in Fig.~\ref{fig:heavyhex_lattices}.
Since this lattice is bipartite, we can write it as
\begin{equation}
    V=V_2 \cup V_3~,\quad E \subset V_2 \times V_3~,
\end{equation}
where $V_i$ consists of vertices with degree $i$.
We further define $W \subset V_2$ as the subset of degree-2 vertices whose two neighbors both belong to $V_3$, and for each $l \in W$ we denote these neighbors by $n_1(l)$ and $n_2(l)$.
Note that $W$ is essentially the bulk qubits of degree 2 in the heavy-hexagonal lattice.

Given $V = \{ 0, 1, \cdots, n-1 \}$ and a vector of spins $\bm{z} = (z_0, \cdots, z_{n-1}) \in \{ \pm 1\}^n$, we define a random spin-glass problem on the heavy-hexagonal lattice with the cost function
\begin{equation}\label{eq:ising_spin_glass}
\begin{split}
    C(\bm{z})=\sum_{v \in V} d_v~z_v+\sum_{(i, j) \in E} d_{i, j}~z_i z_j + \\
    \sum_{l \in W} d_{l, n_1(l), n_2(l)}~z_l z_{n_1(l)} z_{n_2(l)},
\end{split}
\end{equation}
where $d_v$, $d_{i,j}$, and $d_{l, n_1(l), n_2(l)}$ are respectively the linear, quadratic, and cubic coefficients that are sampled uniformly at random from $\{ \pm1 \}$.

We also study Ising spin-glass problems on square lattices with the cost function
\begin{equation}\label{eq:ising_spin_glass_square}
    C(\bm{z}) = \sum_{(i,j)\in E} d_{ij}\, z_i z_j~,
\end{equation}
where $E$ denotes here an edge set reflecting the square-grid connectivity and $d_{ij} \in \{ \pm 1 \}$.
This model contains only nearest-neighbor two-body interactions, in contrast to the heavy-hexagonal case which includes up to three-body terms.

For both problem classes, we let $\mathcal{Z}^\ast = \{\bm{z} : C(\bm{z}) = \min_{\bm{z}'} C(\bm{z}')\}$ denote the set of optimal bitstrings, which may contain multiple elements due to ground-state degeneracy, and write $\bm{z}^\ast \in \mathcal{Z}^\ast$ for any such optimal configuration.
The cost Hamiltonian $H_C$ is obtained by replacing $z_i$ with the Pauli operator $Z_i$ in $C(\bm{z})$.

\section{Method}\label{sec:method}

We now articulate our simulation strategy for QAOA circuits on classical computers.
The primary goal of this paper is to tackle problem instances with large system sizes $n$ using deep circuits, i.e., large $p$.
This regime has remained unexplored mainly due to limitations in both quantum hardware and classical simulation techniques~\cite{QA_vs_QAOA_127, Short_Depth_QAOA_QA, pelofskeScalingWholechipQAOA2024, sachdeva2024quantumoptimizationusing127qubit}.
Regarding quantum hardware, the authors of Ref.~\cite{QA_vs_QAOA_127} benchmarked IBM quantum hardware on the QAOA problem specified by Eq.~\eqref{eq:ising_spin_glass} with $n=127$, using globally optimized parameters obtained via grid search, for circuit depths up to $p=2$.

\subsection{Parameter optimization}\label{subsec:parameter_optimization}

The authors of Ref.~\cite{pelofskeScalingWholechipQAOA2024} showed that Ising spin-glass problems [Eq.~\eqref{eq:ising_spin_glass}] have parameter concentration across problem sizes from $16$ to $127$, which permits transfer learning from small-size problem instances to larger ones.
The parameter concentration has been reported in various problem classes~\cite{Farhi2022quantumapproximate,brandao2018fixedcontrolparametersquantum,boulebnane2022solvingbooleansatisfiabilityproblems}, and is believed to help with the practical use of QAOA for solving large-scale optimization problems.

Based on this observation, we first optimize parameters $\{\bm{\gamma}^{(k)}, \bm{\beta}^{(k)}\}$ for each of $k$ problem instances at a smaller system size $n' < n$, where $n'$ denotes the training system size and $n$ the target system size.
We then compute the median values of the two parameters at each layer across all $k$, and denote the resulting median parameters as $(\bm{\gamma}, \bm{\beta})$, which are applied to the target problem instances with $n$ qubits.

In order to optimize the parameters of deep circuits, we adopt the interpolation-based scheduling technique proposed in Ref.~\cite{apte2025iterativeinterpolationschedulesquantum} in which the parameters are expressed as
\begin{align}
\gamma_j & =\gamma(j / p)=\sum_{c=1}^{\mathcal{C}} u_c f_c(j / p)~, \\
\beta_j & =\beta(j / p)=\sum_{c=1}^{\mathcal{C}} v_c f_c (j / p)~,
\end{align}
where $f_c(x)$ is a family of orthonormal functions on $x \in [0, 1]$ and $x := j/p$ represents the fraction of the schedule.
Additionally, $u_c$ and $v_c$ are the coefficients to be optimized in this scheme, and $\mathcal{C}$ is the number of basis functions chosen such that $\mathcal{C} \ll p$ to restrict the complexity of the optimization process~\cite{apte2025iterativeinterpolationschedulesquantum}.

In this paper, we perform parameter optimization across many problem instances in a small-size training system and take the median parameters as the resulting parameters.
We employ the Chebyshev polynomials rescaled to the interval $[0, 1]$ as the basis functions $f_c(x)$, and optimize the coefficients $u_c$ and $v_c$ using the BOBYQA algorithm~\cite{BOBYQA}.

\subsection{State evolution and sampling methods}\label{subsec:tn_methods}

To simulate these QAOA circuits, we employ a TN ansatz that respects the qubit connectivity of each lattice [see Fig.~\ref{fig:heavyhex_lattices} for heavy-hexagonal lattices].
In our implementation, the TN ansatz consists solely of site tensors, and we rely on the BP algorithm to efficiently perform truncations during gate applications on the TN ansatz~\cite{LEIFER20081899, sahu2022efficienttensornetworksimulation,PhysRevB.108.125111,10.21468/SciPostPhys.15.6.222}.
Specifically, we use converged BP messages as the local environment to the region where a gate is applied and, upon applying the gate, perform a singular value decomposition (SVD) to a fixed bond dimension $\chi$ which maximizes the overlap of the truncated TN with the untruncated TN conditioned on that BP environment~\cite{PhysRevB.99.195105,PhysRevLett.101.090603}.
We exploit the fact that the three-body $Z$ rotation  which appears in the QAOA circuits (Eq.~\eqref{eq:qaoa_ansatz}) for the Hamiltonian in Eq.~\eqref{eq:ising_spin_glass} is decomposable into single-qubit $Z$ rotations and two-qubit $\mathrm{CNOT}$ gates~\cite{QA_vs_QAOA_127}.
In addition, we make use of the BP messages to measure local expectation values directly, paving the way for the optimization of parameters in large-scale systems ($n'$) that remain inaccessible to state-vector simulations [as discussed in Sec.~\ref{subsec:parameter_optimization_TN}].

Once the final state is obtained, we sample bitstrings using the boundary MPS method originally introduced in Ref.~\cite{PhysRevB.104.235141} and extended to arbitrary planar graphs along with corresponding open-source code in Ref.~\cite{rudolph2025simulatingsamplingquantumcircuits}.
Below, we summarize the key idea of this method.
Let $\psi$ denote the TN ansatz. We partition the TN into column-wise components $\psi_b$, where each $\psi_b$ corresponds to column $b = 1, \dots, N_b$ of the planar graph defining the TN topology (while the method can also be applied to row-based partitions, we describe it here in the column-based setting).
We then construct a set of MPSs
$\{ \bm{M}_{N_b \rightarrow N_b-1}, \bm{M}_{N_b-1 \rightarrow N_b-2}, \dots, \bm{M}_{2 \rightarrow 1} \}
$,
where each $\bm{M}_{b \rightarrow b-1}$ approximates the right-hand partition of the norm tensor network $\braket{\psi|\psi}$,
consisting of columns $b$ through $N_b$, i.e.,
$
(\psi_b^\dagger \psi_b)(\psi_{b+1}^\dagger \psi_{b+1}) \cdots (\psi_{N_b}^\dagger \psi_{N_b})
$.
Sampling begins with the leftmost column ($b = 1$), where we compute reduced density matrices from $\psi_1$, $\psi_1^\dagger$, and $\bm{M}_{2 \rightarrow 1}$.
Once a bitstring $x_1$ on the first column is obtained, we construct an MPS denoted $\bm{m}_{1 \rightarrow 2}$ that approximates the projected partition $x_1 \cdot \psi_1$.
For each subsequent column $b > 1$, we repeat the sampling using $\psi_b$ along with the environments $\bm{m}_{b-1 \rightarrow b}$ and $\bm{M}_{b+1 \rightarrow b}$, except for the final column $b = N_b$, where it only requires $\bm{m}_{{N_b}-1 \rightarrow {N_b}}$.

In this paper, we construct two types of MPSs, $\{ \bm{m}_{b-1 \rightarrow b} \}$ and $\{ \bm{M}_{b+1 \rightarrow b} \}$, following the code base of Ref.~\cite{rudolph2025simulatingsamplingquantumcircuits}.
We set the bond dimensions as $R_m$ for the \emph{amplitude} MPSs $\{ \bm{m}_{b-1 \rightarrow b} \}$ and $R_M$ for the \emph{norm} MPSs $\{ \bm{M}_{b+1 \rightarrow b} \}$.
In the limit $R_m, R_M \to \infty$, these MPSs yield exact partial contractions of the network.
We then define the target distribution $P(\bm{z}) := |\!\braket{\bm{z}|\psi}\!|^2$ and the sampled distribution $Q(\bm{z})$ obtained via the boundary MPS sampling procedure.
By construction, $P(\bm{z})$ is determined solely by the accuracy of the amplitude MPSs with bond dimension $R_m$,
whereas $Q(\bm{z})$ depends on both the amplitude and norm MPSs; in practice, bringing $Q(\bm{z})$ close to $P(\bm{z})$ is largely governed by the choice of the bond dimension $R_M$.
Under these finite bond dimensions, the procedure effectively implements an importance-sampling scheme~\cite{review_of_is}.
We thereby define the importance weight
\begin{equation}
    \omega := \frac{P(\bm{z})}{Q(\bm{z})}~,
    \label{eq:importance_weight}
\end{equation}
which quantifies how well $Q(\bm{z})$ approximates $P(\bm{z})$, and this is used to assess the reliability of our sampling procedure.

\section{Results}\label{sec:results}

\subsection{Benchmarks on heavy-hexagonal lattice instances}\label{subsec:benchmarks_heavyhexagonal}

We first analyze the performance of different circuit depths on a heavy-hexagonal problem instance [Eq.~\eqref{eq:ising_spin_glass}] defined on the \texttt{ibm\_washington} architecture with $n = 127$ qubits [see Fig.~\ref{fig:heavyhex_lattices}(d)].
We obtain circuit schedules for $p \in \{ 10, 50, 100 \}$ by optimizing the parameters via an interpolation schedule at $\mathcal{C}=10$ on $100$ training instances of size $n'=16$ [see Fig.~\ref{fig:heavyhex_lattices}(a)], taking the median parameters across instances. We use  exact state-vector simulations for the training, as $n'$ is small enough.
For the TN simulations, we adopt $\chi = 32$ as the bond dimension for the BP approximation, and set the amplitude rank $R_m = \chi$ and the norm rank $R_M = 1$ for the boundary MPS sampling.

\begin{figure}[t]
    \centering
    \includegraphics[clip, width=3.0in]{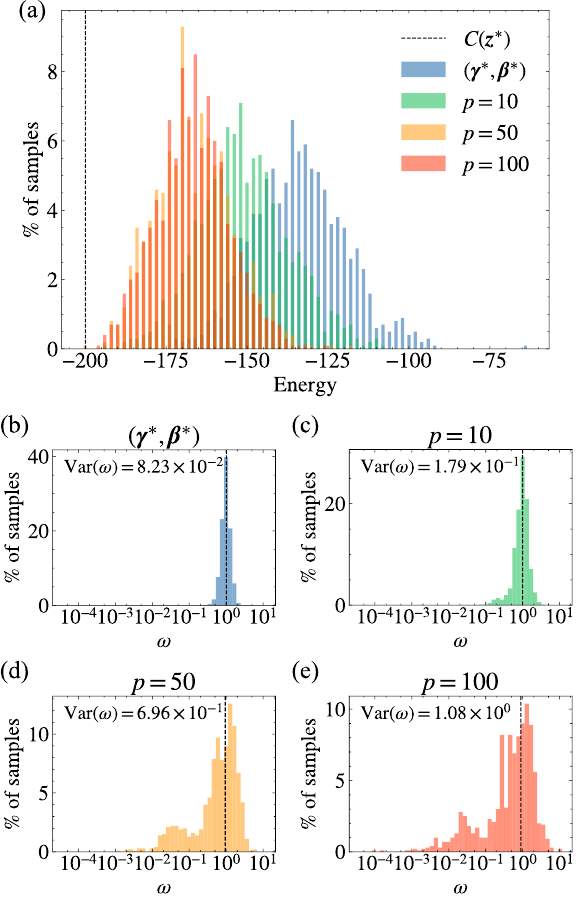}
    \caption{
    (a) Histogram of sampled energies for the $127$-qubit 0th \texttt{ibm\_washington} instance defined in Ref.~\cite{QA_vs_QAOA_127}, where the minimum energy is $-200$ denoted by the vertical dashed line.
    Results are shown for the baseline parameters $(\bm{\gamma}^\ast, \bm{\beta}^\ast)$ with $p = 5$ [Eq.~\eqref{eq:p5_parameters}] and for deeper schedules with $p = 10$, $50$, and $100$.
    (b)--(e) Distributions of importance sampling weights $\omega$ for $(\bm{\gamma}^\ast, \bm{\beta}^\ast)$, $p=10$, $50$, and $100$, respectively; the mean values of the weights are illustrated by vertical dashed lines, and the variances are indicated in the upper-left corner of each panel.
    The data are obtained with bond dimension $\chi = 32$, amplitude-MPS rank $R_m = \chi$, and norm-MPS rank $R_M = 1$.
    }
    \label{fig:ibm_washington_0}
\end{figure}

Fig.~\ref{fig:ibm_washington_0}~(a) plots the sampled energies for the 0th instance of \texttt{ibm\_washington} computed from $1{,}000$ bitstrings.
As a baseline, we also include the results using the parameters
\begin{align}\label{eq:p5_parameters}
\bm{\gamma}^{\ast} &= [6.16555, 6.08373, 6.01445, 5.9616, 5.93736]~,\\
\bm{\beta}^{\ast} &= [0.53822, 0.44776, 0.32923, 0.23056, 0.12587]~,
\end{align}
which were obtained by a global parameter search at $p=5$ in Ref.~\cite{pelofskeScalingWholechipQAOA2024}.
The results clearly demonstrate that deeper schedules tend to yield better solutions on average.
However, no drastic improvement is observed for $p \ge 50$, indicating that the parameter transfer from $n'=16$ has saturated at this depth.

The corresponding sampling weights $\omega$ obtained from different circuit schedules are shown in Figs.~\ref{fig:ibm_washington_0}~(b)--(e).
The mean values of the weights, illustrated by vertical dashed lines, are close to unity, implying that the norm of the TN state is well preserved under BP truncations.
As can be seen from Figs.~\ref{fig:ibm_washington_0}~(b) and (c), the sampling weights have small variances, denoted by $\mathrm{Var}(\omega)$, demonstrating that the samples are sufficiently accurate even with such a small norm rank $R_M = 1$.
On the other hand, Figs.~\ref{fig:ibm_washington_0}~(d) and (e), corresponding to the results from schedules with $p \in \{ 50, 100 \}$, exhibit significantly larger variances.

\begin{figure}
    \centering
    \includegraphics[clip, width=3.0in]{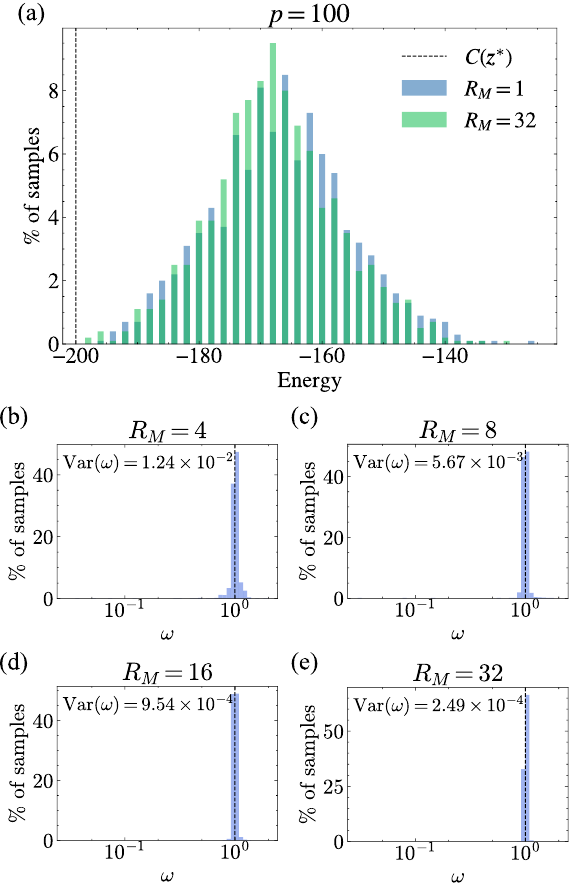}
    \caption{Dependence on the norm-MPS rank $R_M$ for the same instance as in Fig.~\ref{fig:ibm_washington_0} at $p = 100$.
    (a) Histogram of sampled energies for $R_M = 1$ and $32$.
    (b)--(e) Distributions of importance sampling weights $\omega$ for $R_M = 4$, $8$, $16$, and $32$, respectively.}
    \label{fig:ibm_washington_0_variousR}
\end{figure}

To examine how $R_M$ influences the quality of importance sampling, we carry out sampling for the state at $p = 100$ using different norm MPS ranks $R_M \in \{ 4, 8, 16, 32 \}$.
Fig.~\ref{fig:ibm_washington_0_variousR}(a) compares the energy distributions for $R_M = 1$ and $R_M = 32$, showing a slight shift toward lower energies with increasing $R_M$ but no major overall change.
In contrast, Figs.~\ref{fig:ibm_washington_0_variousR}(b)--(e) demonstrate that increasing $R_M$ markedly improves sampling quality; the variance shrinks significantly as $R_M$ increases.
These results interestingly indicate that small $R_M$ values are still sufficient to obtain low-energy samples, as the overall energy distribution remains largely unchanged between $R_M = 1$ and $R_M = 32$, despite significant improvement in the accuracy of the samples themselves.

To characterize the complexity of the quantum states during the circuit evolution, we analyze the entanglement behavior of the TN ansatz.
The BP algorithm, upon convergence, provides two messages on the edge $(i,j) \in E$, $M_{i,j}$ and $M_{j,i}$, where $M_{i,j} := \left(M^{1/2}_{i,j}\right)^{\dagger}\left(M^{1/2}_{i,j}\right)$ because the message tensors remain positive semi-definite throughout the BP algorithm as long as they are initialized as such.
We can obtain entanglement properties of the state via the singular values of the product $M_{i,j}^{1/2} M_{j,i}^{1/2}$ which is a $\chi \times \chi$ matrix.
Specifically, we can define an effective Von Neumann entanglement entropy for the edge $(i,j)$ as
\begin{equation}\label{eq:entanglement_entropy}
\mathcal{S}_{i,j} = - \sum_{\ell} [\Lambda^2_{i,j}]_{\ell} \log_{2} [\Lambda^2_{i,j}]_{\ell} ~,
\end{equation}
where $[\Lambda_{i,j}]_{\ell}$ are the singular values of the product $M_{i,j}^{1/2} M_{j,i}^{1/2}$ obtained by SVD.
In our gate evolution, the norm of the TN ansatz is not exactly preserved, so we normalize $[\Lambda_{i,j}]_{\ell}$ to ensure the entanglement is taken with respect to a state with $\langle \psi \vert \psi \rangle \approx 1$.
For a large region $\mathcal{A}$ of the tensor network with a boundary of edges $\partial \mathcal{A}$ separating it from the rest of the TN, the BP approximation to the entanglement entropy of that region is (see Fig.~\ref{fig:ibm_geneva_0_entanglements} for illustration here)
\begin{equation}
    S_{\mathcal{A}} = \sum_{(i,j) \in \partial \mathcal{A}}S_{i,j}~.
\label{eq:Scut}
\end{equation}
In a lattice with large loops such as the heavy-hex lattice, and for a sufficiently large bipartition, $S_{\mathcal{A}}$ will be a very good approximation to the entanglement entropy of that region.

\begin{figure}[tbp]
    \centering
    \includegraphics[clip, width=3.0in]{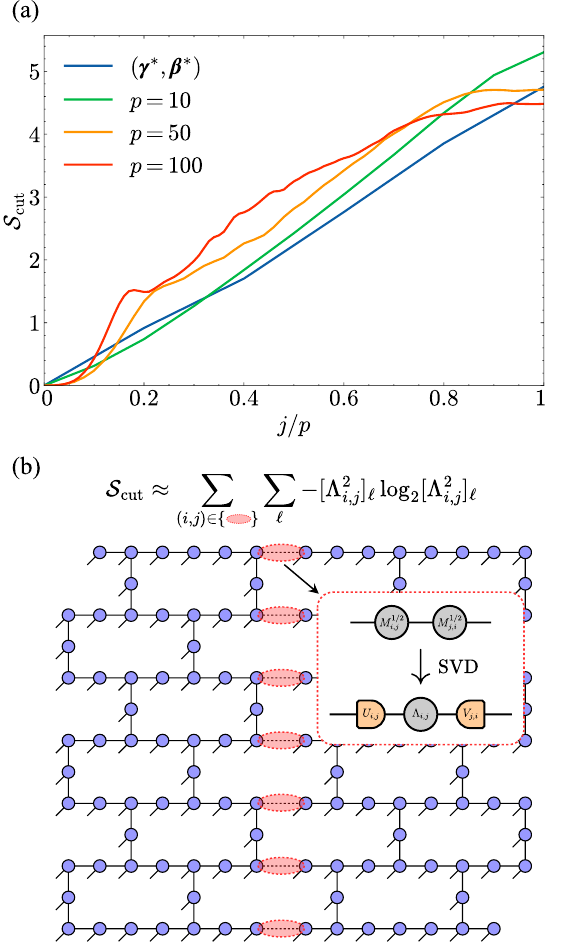}
    \caption{%
    Bipartite entanglement behavior during the QAOA simulations for the same instance as in Fig.~\ref{fig:ibm_washington_0}.
    (a) The bipartite, Von-Neumann entanglement entropy $\mathcal{S}_{\mathrm{cut}}$, approximated by Eq.~\eqref{eq:Scut}, across edges bisecting the TN state [see (b)], plotted against the normalized circuit step $j/p$.
    (b) Schematic illustration of the approximate derivation of $\mathcal{S}_{\mathrm{cut}}$, where dashed lines (circled by red ovals) indicate the set of edges $\partial \mathcal{A}$.
    The effective von Neumann entanglement entropy [Eq.~\eqref{eq:entanglement_entropy}] for each edge is read off by the (converged) BP messages residing on an edge of TN ansatz as depicted in the inset.
    Given the bond dimension $\chi = 32$ used in the simulation, the maximum attainable value of $\mathcal{S}_{\mathrm{cut}}$ for our TN ansatz is $35$.
    }
    \label{fig:ibm_washington_entanglements}
\end{figure}

Figure~\ref{fig:ibm_washington_entanglements}(a) presents the values of $S_{\mathrm{cut}}$ defined by Eq.~\eqref{eq:Scut} at each step, where $\partial \mathcal{A}$ is chosen in $S_{\mathrm{cut}}$ to vertically bisect the TN state as shown in Fig.~\ref{fig:ibm_washington_entanglements}(b).
As shown in Fig.~\ref{fig:ibm_washington_entanglements}(a), shallower schedules with $p \le 10$ exhibit linear growth in $\mathcal{S}_{\mathrm{cut}}$.
On the other hand, for deeper schedules, $\mathcal{S}_{\mathrm{cut}}$ saturates and ceases to increase further.
Furthermore, we observe that the peak values of $\mathcal{S}_{\mathrm{cut}}$ remain lower than those with shallow schedules.
These results indicate that our TN ansatz performs effectively for simulating the QAOA circuits, even, or rather especially with deep schedules, provided that well-optimized parameters are available.
Note that since the theoretical upper bound of Eq.~\eqref{eq:entanglement_entropy} is given by $\log_2 \chi$, the observed entanglement $\mathcal{S}_{\mathrm{cut}}$ reaches values significantly smaller than the limit of $7\log_2 32 = 35$ for a fixed bond dimension $\chi = 32$.

Consequently, while parameters transferred from $n'=16$ do not by themselves guarantee reaching the global optimum for the $n=127$ instances, the simulations consistently produce low-energy samples and exhibit modest entanglement growth, which is preferable for TN simulations at a fixed bond dimension.
In contrast, as presented in Appendix~\ref{appendix:ibm_geneva}, the parameter transfer performs remarkably well for $n=27$-qubit instances defined on \texttt{ibm\_geneva} [Fig.~\ref{fig:heavyhex_lattices}(b)], where circuits with $p \ge 50$ frequently yield exact solutions.
These observations provide empirical evidence that achieving the global optimum in small training systems may hinder the generalization performance when scaled to larger target systems.

\subsection{Parameter optimization based on TN simulations}\label{subsec:parameter_optimization_TN}

\begin{figure*}[tbp]
    \centering
    \includegraphics[clip,width=0.95\textwidth]{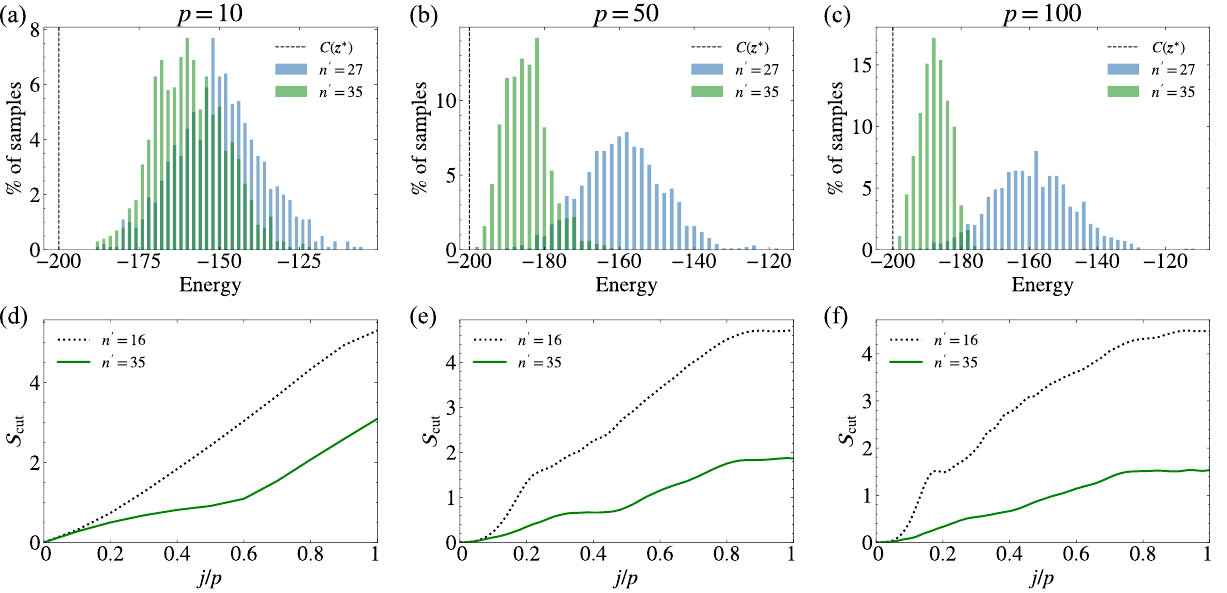}
    \caption{%
    (a)--(c) Histograms of sampled energies for the $127$-qubit 0th \texttt{ibm\_washington} instance at three circuit depths $p\in \{ 10, 50, 100\}$, comparing two parameter sets optimized on training systems with $n' = 27$ and $35$, respectively.
    Lattices defining the training instances with $n'=27$ and $n'=35$ are illustrated in Figs.~\ref{fig:heavyhex_lattices}(b) and (c).
    (d)--(f) Bipartite entanglement behavior during QAOA simulations using parameters trained on the $n'=35$ system (solid green lines), where $\mathcal{S}_{\mathrm{cut}}$ is defined in the caption of Fig.~\ref{fig:ibm_washington_entanglements}. 
    For comparison, the $n'=16$-trained results from Fig.~\ref{fig:ibm_washington_entanglements}(a) are overlaid as dotted lines.
}
    \label{fig:ibm_washington_0_instance_set}
\end{figure*}

Having observed that parameters optimized on small training instances do not generalize well to larger heavy-hexagonal systems [see Sec.~\ref{subsec:benchmarks_heavyhexagonal}], we next attempt to transfer parameters from larger training systems.
Since system sizes of $n' \gtrsim 30$ are generally beyond the reach of efficient state-vector simulations, we adopt TN methods for the optimization.
We intentionally opt for a relatively small bond dimension, $\chi = 8$, to accommodate the repeated state evolutions required during the BOBYQA algorithm.
In this process, the expectation values of the cost Hamiltonian [Eq.~\eqref{eq:ising_spin_glass}] are computed by contracting local spin operators with the TN ansatz and its conjugate.
This contraction is efficiently performed using the BP message tensors~\cite{10.21468/SciPostPhys.15.6.222} as mentioned in Sec.~\ref{subsec:tn_methods}.

Figures~\ref{fig:ibm_washington_0_instance_set}(a)--(c) present a comparison of the results obtained using $n'=27$ and $n'=35$ training systems illustrated in Figs.~\ref{fig:heavyhex_lattices}(b) and (c).
The energy distributions for circuit schedules with $p=10, 50$, and $100$ are shown, each based on $1{,}000$ samples.
Here, we compute the median parameters from ten problem instances for each training size, rather than the one hundred used in the previous section.
We observe that the results from $n' = 27$ training systems exhibit nearly identical distributions to those from $n' = 16$ training systems [Fig.~\ref{fig:ibm_washington_0}].
This result is consistent with our observations in Appendix~\ref{appendix:ibm_geneva}, where the $n'=16$ parameters already yield exact solutions for the $n=27$ instances.
The lack of further improvement suggests that the $n'=27$ landscape provides insufficient gradient to further refine the parameters for larger-scale generalization.
We note that despite the fact that both results remain trapped in nearly the same mode, the energy at $p=100$ shows a wider distribution compared to $p=50$.
We surmise that this occurs because approximate TN optimization generally yields different schedules compared to state-vector simulations; these discrepancies likely become more pronounced as the circuit depth $p$ increases.

In contrast, parameters trained at $n'=35$ yield a significant improvement in the sampled energies on average.
As shown in Fig.~\ref{fig:ibm_washington_0_instance_set}(a), this improvement is already apparent at shallow depths ($p=10$).
Figures~\ref{fig:ibm_washington_0_instance_set}(b) and (c) further show that the optimization at $n'=35$ successfully escapes the local minima encountered with $n' \in \{16, 27\}$ training systems, making steady progress toward better solutions.
In summary, parameters optimized on $n'=35$ instances exhibit superior generalization to unseen larger systems compared to those derived from small-size training systems.
Notably, this energy improvement is accompanied by a reduction in the bipartite entanglement entropy $\mathcal{S}_{\mathrm{cut}}$ [Figs.~\ref{fig:ibm_washington_0_instance_set}(d)--(f)], keeping the resulting circuits well-suited for TN simulation.

\subsection{Benchmarks on square lattice instances}\label{subsec:benchmarks_square_lattices}

\begin{figure*}[htbp]
    \centering
\includegraphics[clip,width=0.95\textwidth]{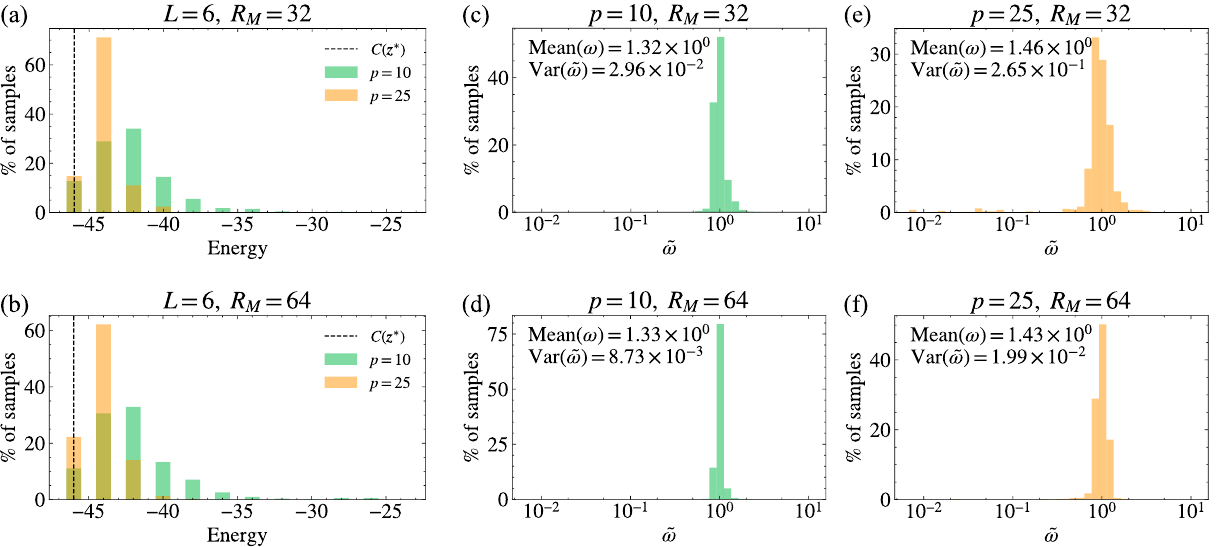}
    \caption{%
    Sampling results for a $6\times6$ square-lattice Ising spin-glass instance whose ground-state energy is $C(\bm{z}^\ast) = -46$.
    (a) and (b) show histograms of the sampled energies obtained with norm-MPS ranks $R_M = 32$ and $64$, respectively, each at circuit depths $p = 10$ and $25$.
    (c, d) and (e, f) show the corresponding distributions of $\tilde{\omega}$; $\mathrm{Mean}(\omega)$ and $\mathrm{Var}(\tilde{\omega})$ are indicated in each panel.}
    \label{fig:square_lattice_results_L=6}
\end{figure*}

\begin{figure*}[htbp]
    \centering
\includegraphics[clip,width=0.95\textwidth]{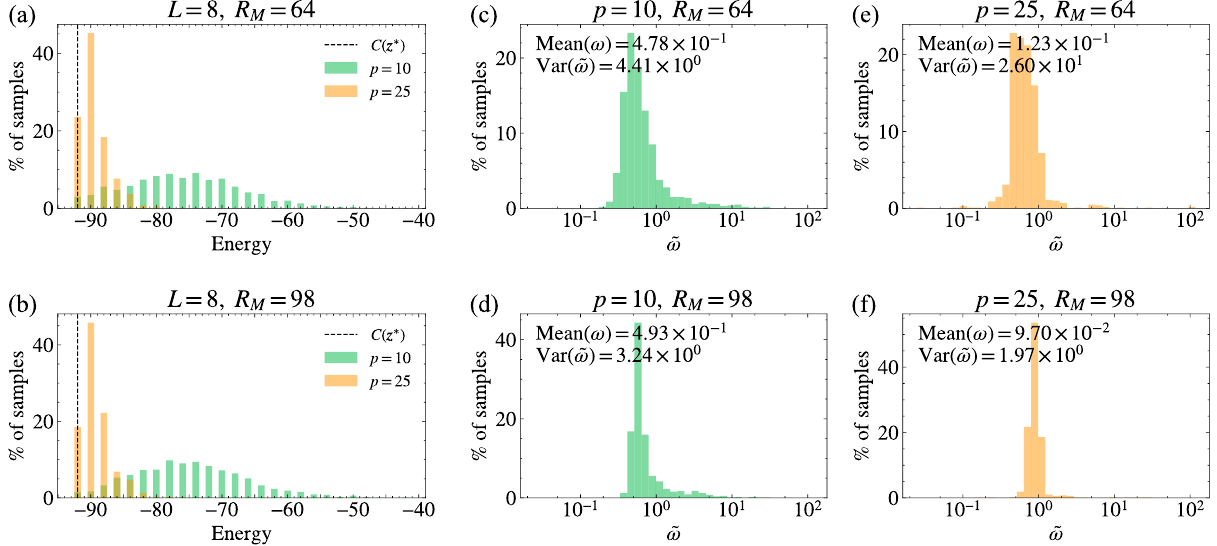}
    \caption{%
    Sampling results for a $8\times8$ square-lattice Ising spin-glass instance whose ground-state energy is $C(\bm{z}^\ast) = -92$.
    (a) and (b) show histograms of the sampled energies obtained with norm-MPS ranks $R_M = 64$ and $98$, respectively, each at circuit depths $p = 10$ and $25$.
    (c, d) and (e, f) show the corresponding distributions of $\tilde{\omega}$; $\mathrm{Mean}(\omega)$ and $\mathrm{Var}(\tilde{\omega})$ are indicated in each panel.}
    \label{fig:square_lattice_results_L=8}
\end{figure*}

We next study the case for square-lattice Ising spin-glass problems [Eq.~\eqref{eq:ising_spin_glass_square}].
We perform the parameter optimization across $100$ problem instances defined on a $4\times4$ grid lattice (i.e. $n'=16$) using state-vector simulations, with the same optimization settings as in Sec.~\ref{subsec:benchmarks_heavyhexagonal}, and transfer the parameter to $6\times6$ ($n=36$) and $8\times8$ ($n=64$) square-lattices.
When simulating the QAOA circuit for these lattices, we adopt a TN ansatz that respects their connectivity and thus the connectivity induced by the circuits.

To analyze the sampling quality, let us define the normalized importance weights $\tilde{\omega}$.
Recall the importance weights $\omega = P(\boldsymbol{z})/Q(\boldsymbol{z})$ defined in Eq.~\eqref{eq:importance_weight}.
The mean of $\omega$ is in general governed by the norm of the TN state through the relation
\begin{equation}
    \mathbb{E}_{\boldsymbol{z} \sim Q}[\omega(\boldsymbol{z})]
    = \sum_{\boldsymbol{z}} P(\boldsymbol{z})
    = \braket{\psi|\psi}~,
\end{equation}
where $\ket{\psi}$ is a TN state that does not necessarily have unit norm as it is not preserved during the truncations following gate applications.
Whilst for heavy-hexagonal lattices we find the mean of $\omega$ remains close to unity (Sec.~\ref{subsec:benchmarks_heavyhexagonal}), the deviation from unity is pronounced in the square-lattice simulations~\cite{rudolph2025simulatingsamplingquantumcircuits}, where the BP approximation is less reliable due to short loops that undermine the factorization of the local environment into independent BP messages.
We therefore define the normalized importance weights as
\begin{equation}
    \tilde{\omega}
    =
    \frac{\omega}{
    \frac{1}{N_s}\sum_{i=1}^{N_s} \omega_i
    }~,
\end{equation}
where $N_s$ is the number of samples and $\{\omega_i\}_{i=1}^{N_s}$ are the sampled importance weights.
That is, $\omega$ is normalized by its arithmetic mean ($\mathrm{Mean}(\omega)$) so that the distribution is centered at unity.
When $N_s$ is large enough for the sample mean to converge to the TN norm $\braket{\psi|\psi}$, the variance of $\tilde{\omega}$ serves as a meaningful diagnostic of the sampling accuracy: a small variance then implies that $\omega$ is concentrated near $\braket{\psi|\psi}$, which is the signature of a faithful sampler.
It is worth noting that $\mathrm{Var}(\tilde{\omega})$ alone is not a sufficient diagnostic without this convergence condition: a biased sampler that consistently returns configurations with a nearly constant---but incorrect---weight would still yield a narrow $\tilde{\omega}$ distribution.

In our TN simulations, we set the bond dimension to $\chi=16$, smaller than the value used in Sec.~\ref{subsec:benchmarks_heavyhexagonal}, to allow us to use sufficiently large boundary-MPS ranks relative to $\chi$.
These choices are not always justified in terms of optimizing the simulation accuracy, but in practice we find they can be remarkably effective for efficiently obtaining reasonable solutions via sampling.
When sampling from the resulting TN states, we set the amplitude-MPS rank to $R_m = 2\chi$ and vary the norm-MPS rank $R_M$ depending on the system size (see the captions of Figs.~\ref{fig:square_lattice_results_L=6} and~\ref{fig:square_lattice_results_L=8}); we take $1,000$ samples for each case.
As a technical note, we utilize GPU acceleration for the importance sampling \cite{rudolph2025simulatingsamplingquantumcircuits}, as boundary MPS contractions on square-lattice tensor networks are computationally demanding.

Figures~\ref{fig:square_lattice_results_L=6} and~\ref{fig:square_lattice_results_L=8} present the results for representative $6\times6$ and $8\times8$ instances with ground-state energies $C(\bm{z}^\ast) = -46$ and $-92$ respectively.
We confirm that parameter concentration remains effective for these models: as $p$ increases, lower energies are consistently obtained [panels~(a) and~(b)].
Turning to the sampling quality, panels~(c, d) and~(e, f) show the distributions of $\tilde{\omega}$ for the two values of $R_M$.
In principle, one should verify that the sample mean of $\omega$ has converged to $\braket{\psi|\psi}$ before relying on $\mathrm{Var}(\tilde{\omega})$.
Since computing $\braket{\psi|\psi}$ directly is, however, expensive for the square-lattice TN states considered here, we instead monitor the stability of $\mathrm{Mean}(\omega)$ across $R_M$ as a practical indicator.
For the $6\times6$ instance, $\mathrm{Mean}(\omega)$ is already converged at both values of $R_M$, while $\mathrm{Var}(\tilde{\omega})$ decreases with $R_M$, validating the sampling accuracy.
For the $8\times8$ instance, the stabilization of $\mathrm{Mean}(\omega)$ cannot be seen within the range of $R_M$ that we can afford; nevertheless, the mode of $\tilde{\omega}$ approaches unity as $R_M$ increases, and we therefore rely on $\mathrm{Var}(\tilde{\omega})$ and observe that it decreases with $R_M$.

\section{Conclusion}\label{sec:conclusion}

We presented a TN framework for simulating QAOA on planar architectures using a connectivity-aware TN ansatz, where truncations during gate application are performed under
the BP approximation and samples are generated via the boundary MPS algorithm.
We target Ising spin-glass problems defined on IBM's heavy-hex and square lattice architecture, and evaluate the impact of parameter concentration for deep circuit schedules.
We also extend the training system sizes beyond the reach of state-vector simulations by using TN simulations.

First, transferring parameters from systems an order of magnitude smaller improves with schedule depth but saturates beyond the depth at which the smaller training systems already reach optimality; this clearly delineates a practical limit of parameter transfer.
Second, parameter training on representative system sizes mitigates the inherent limitations of parameter concentration; this approach remains effective even when coarse optimizations via TN simulations introduce residual inaccuracies.
Finally, we demonstrate that our TN approach is effective for identifying parameter concentration in square-lattice instances.
Although this model requires significantly greater computational resources, GPUs serve as a powerful workhorse to meet these demands.

Our observations are consistent with Ref.~\cite{pelofske2025evaluatinglimitsqaoaparameter}, which likewise reports for heavy-hexagonal instances that parameter concentration works effectively for deep circuit schedules and that TN ans\"{a}tze are well-suited to simulate QAOA circuits of locally interacting problems.
Going beyond these findings, we further analyzed the entanglement growth and the sampling quality. In particular, we show that as long as the parameter schedules are well-optimized, even deep circuits of $p=100$ layers — a regime not explored in prior work — remain numerically tractable within our TN framework. Furthermore, we demonstrated that parameter optimization via TN simulations can facilitate solving large-scale problems, extending the analysis to square lattices which were not considered in Ref.~\cite{pelofske2025evaluatinglimitsqaoaparameter}.

Taken together, our results indicate that TN methods, combined with a sampling technique on two-dimensional architectures, provide a practical analysis tool for variational quantum algorithms on modern hardware and are well-suited for benchmarking deep schedules on systems with hundreds of qubits.
They also open a path to performing parameter optimization with classical surrogates~\cite{rudolph2023classicalsurrogatesimulationquantum,lerch2024efficientquantumenhancedclassicalsimulation}.

More broadly, these findings underscore the importance of algorithm design in arguing for quantum utility.
In particular, a spatially local spin-glass Hamiltonian can be addressed efficiently and accurately using TN methods via imaginary-time evolution or other spectral filters~\cite{watanabe2026quantuminspiredalgorithmclassicalspin}, and numerous strong classical algorithms exist for broad types of combinatorial optimization problems~\cite{Fan2023}.
Accordingly, it is important to identify problem families for which QAOA is genuinely favorable, and to explore alternative models of quantum computation beyond the settings considered here.

\section*{acknowledgments}

The results in this paper were obtained using the \textit{TensorNetworkQuantumSimulator.jl} library \cite{rudolph2025simulatingsamplingquantumcircuits}, an open-source Julia package built on top of ITensors.jl \cite{itensor} for optimizing and contracting tensor networks of arbitrary topology.
This work was supported by JST ASPIRE Grant Number JPMJAP2319 and by AFOSR under Award No. FA9550-21-1-0236.
R.W. also acknowledges support from JSPS KAKENHI No. 25KJ1773, MEXT Q-LEAP No. JPMXS0120319794, and JST CREST No. JPMJCR24I1.
The authors are grateful for the ongoing support from the Flatiron Institute, a division of the Simons Foundation.

\bibliography{main}

\appendix

\section{Benchmark for \texttt{ibm\_geneva} instances}\label{appendix:ibm_geneva}

\begin{figure}[tbp]
    \centering
    \includegraphics[clip,width=3.0in]{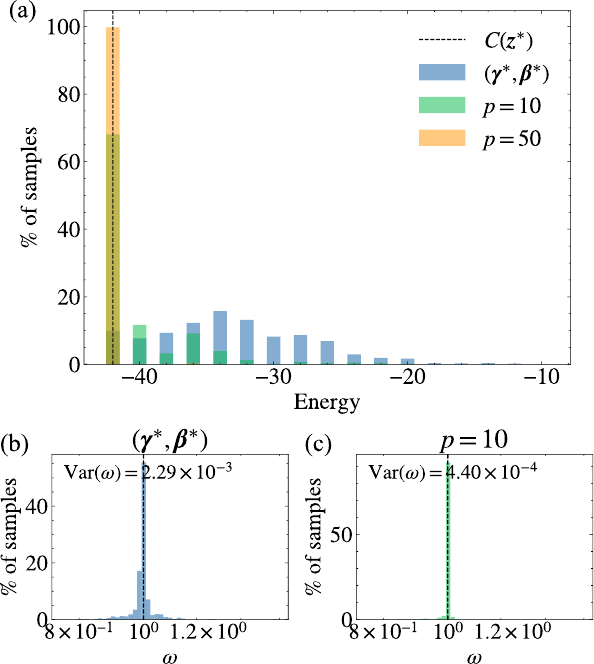}
    \caption{
    (a) Histogram of sampled energies for the $27$-qubit 0th \texttt{ibm\_geneva} instance in Ref.~\cite{QA_vs_QAOA_127}, where the minimum energy is $C(\bm{z}^\ast) = -42$.
    The labels are the same as in Fig.~\ref{fig:ibm_washington_0}(a).
    (b)-(c) Distributions of importance sampling weights $\omega$ corresponding to the labels $(\bm{\gamma}^\ast, \bm{\beta}^\ast)$ and $p=10$ schedules, respectively. At $p=50$, we obtained $\mathrm{Var}(\omega) = 3.56 \times 10^{-8}$.}
    \label{fig:ibm_geneva_0}
\end{figure}

We here discuss the results for an instance of \texttt{ibm\_geneva} [Fig.~\ref{fig:heavyhex_lattices}(b)] following the same settings as in Sec.~\ref{subsec:benchmarks_heavyhexagonal}, where we transferred the parameters optimized on \texttt{ibmq\_guadalupe} instances.
Fig.~\ref{fig:ibm_geneva_0} plots the energy histogram for the 0th instance of \texttt{ibm\_geneva} defined in Ref.~\cite{QA_vs_QAOA_127}, obtained from circuits using the parameters in Eq.~\eqref{eq:p5_parameters} and the optimized parameters with $p \in \{ 10, 50 \}$.
From Fig.~\ref{fig:ibm_geneva_0}(a), we observe that the optimal solution $C(\bm{z}^\ast) = -42$ is obtained in most cases even within $p = 10$, and the optimal solution is fully recovered at $p = 50$.
Figs.~\ref{fig:ibm_geneva_0}~(b) and (c) show that the sampling weights are tightly concentrated around $\omega = 1$ with negligible variance; for instance, at $p = 50$ we obtain $\mathrm{Var}(\omega) = 3.56 \times 10^{-8}$, indicating that the sampling is nearly exact.

These results indicate clearly that the parameter transfer from $n' = 16$ training systems has already saturated by $p = 50$, and no further improvement is obtained beyond this depth.
This may be the reason that these parameters lose their generalization performance for large-size instances with $n=127$ as discussed in Sec.~\ref{subsec:benchmarks_heavyhexagonal}.

\begin{figure}[t]
    \centering
    \includegraphics[clip,width=3.0in]{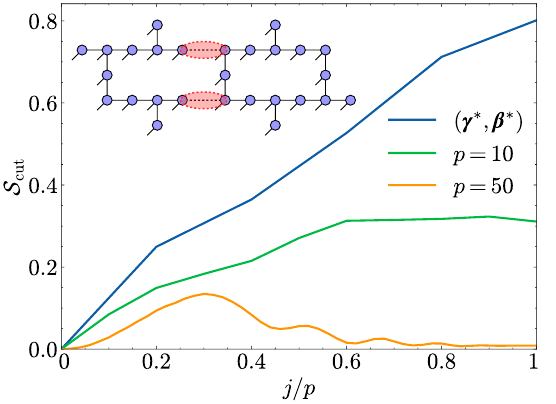}
    \caption{
    Approximate bipartite entanglement entropy $\mathcal{S}_{\mathrm{cut}}$ computed as the sum of Eq.~\eqref{eq:entanglement_entropy} across the bonds corresponding to the edges illustrated by the dashed lines in the inset. The inset shows our TN ansatz that respects the connectivity of \texttt{ibm\_geneva} architecture.}
    \label{fig:ibm_geneva_0_entanglements}
\end{figure}

We examine the entanglement behavior for this instance, as shown in Fig.~\ref{fig:ibm_geneva_0_entanglements}(a).
The entanglement entropy $\mathcal{S}_{\mathrm{cut}}$ is defined as the sum of Eq.~\eqref{eq:entanglement_entropy} across the bonds corresponding to the edges illustrated by the dashed lines in Fig.~\ref{fig:ibm_geneva_0_entanglements}(b).

The entanglement for the $p=50$ schedule is found to converge to nearly zero, and slight entanglement peaks appear at intermediate steps of the schedule, while the $p=10$ schedule exhibits a nonzero entanglement plateau after the peak.
This suggests that numerical truncations in BP approximations can drive the state toward a specific configuration, causing the entanglement to vanish, even when the cost Hamiltonian in Eq.~\eqref{eq:ising_spin_glass} possesses degenerate ground states~\cite{watanabe2026quantuminspiredalgorithmclassicalspin}.

The schedule of $(\bm{\gamma}^{\ast}, \bm{\beta}^{\ast})$ (blue line) exhibits a monotonic increase in $\mathcal{S}_{\mathrm{cut}}$, reaching values significantly higher than the peaks observed in any other schedules.
These results highlight that our TN approach is particularly well-suited for simulating QAOA circuits along the ``optimal'' parameter paths, where the suppressed entanglement growth ensures both numerical stability and computational efficiency.
This observation is consistent with the results discussed in Sec.~\ref{subsec:benchmarks_heavyhexagonal}, but provides an even more compelling argument.

\end{document}